# Raman spectroscopy of chocolate bloom

Siyu He[1,2] and Dmitri V. Voronine[1,3]

[1] Texas A&M University, College Station, TX 77843, USA
[2] Xi'an Jiaotong University, Xi'an 710049, China
[3] Baylor University, Waco, TX 76798, USA

**Abstract**

Raman spectroscopy has been widely applied to the chemical analysis of food quality and, in particular, of chocolate. We perform Raman analysis of bloom formation in white and dark chocolates. We show evidence of sugar and fat bloom and investigate spectroscopic signatures of polymorphic changes in different phases.

1. **Introduction**

Rapid, reliable determination of the distribution and microstructure of the ingredients is crucial in the food industry for keeping the high standards of the quality and safety of food. Chocolate bloom is a characteristic whitish coating appearing on the surface as a sign of poor quality. Two types of bloom, namely fat and sugar bloom, may occur due to fast temperature changes, improper storage and other reasons.[1,2] Previous studies used x-ray photoelectron spectroscopy, environmental scanning electron microscopy, and other techniques to detect and characterize chocolate bloom. Here we obtain spectroscopic Raman signatures of both types of bloom.

Raman spectroscopy, has been widely applied to the chemical analysis of food[3,4,5]. It provides rich information about molecular structure, conformation, crystallization, etc[6]. We used Raman spectroscopy to characterize bloomed chocolate and revealed the chemical distribution changes caused by the bloom. Our approach may be used for rapid label-free analysis of the properties and quality of chocolate.

We focused the analysis on white chocolate to reveal the characteristic Raman bands of different states whereas the analysis of dark chocolate is more challenging due to the fluorescence of cocoa liquor. Nevertheless, we were also able to obtain Raman spectra of dark chocolate and compare with white chocolate. Dark chocolate mostly consists of a mixture of cocoa butter and sugar, while white chocolate is a confection based on sugar, milk and cocoa butter without cocoa liquor[7]. We investigated the ability of Raman spectroscopy to distinguish between sugar and cocoa butter in different states of dark and white chocolate.

2. **Results and discussion**

The samples of white and dark chocolate were obtained from a local store (Hershey brand). The liquid chocolate bloomed after storage at ~ 28 °C for two days. These procedure was used to obtain the liquid and bloomed chocolates. Fig. 1 shows photographs of the white and dark chocolate samples including solid white (SW), solid dark (SD), liquid white (LW), liquid dark (LD), bloomed white (BW), and bloomed dark (BD). Two excitation laser wavelengths at 532 nm and 785 nm and the 1200 gr $mm^{-1}$ grating were used. LabRAM Raman microscope (Horiba) was used for measuring Raman spectra.

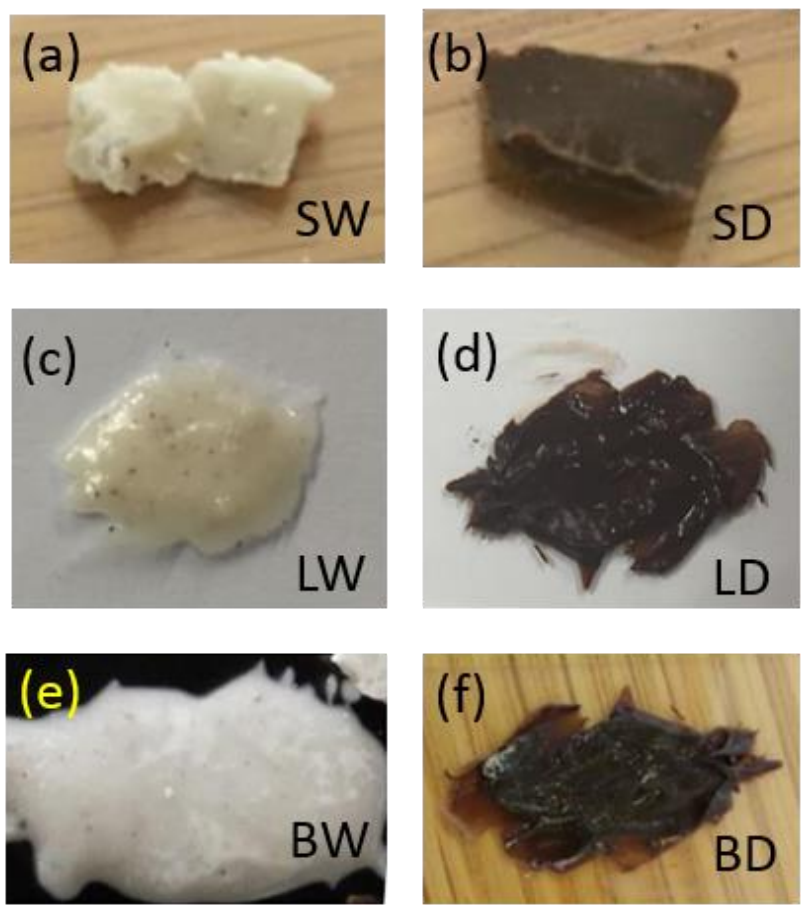

**Figure 1.** Photographs of white and dark chocolate: (a) solid white (SW), (b) solid dark (SD), (c) liquid white (LW), (d) liquid dark (LD), (e) bloomed white (BW), and (f) bloomed dark (BD).

Cocoa butter consists of triacylglycerols and has six polymorphs, labeled in order of increasing melting point[5,6,7]. Form $\beta$(V) is the main polymorph in well-tempered chocolate, solid at room temperature and liquid at body temperature, satisfying consumer standards.[2] However, $\beta$(VI) is the most stable polymorph of cocoa butter as it has the densest packing of triglyceride molecules that comprise the fat[2]. Therefore $\beta$(VI) is considered as the main reason of the fat chocolate bloom[2,8]. Recently, polymorphs were studied using Raman spectroscopy, which revealed spectral signatures of various molecular functional groups.[6] Lactose and sucrose are the main sugars which are abundant in white and less abundant in dark chocolate. Typical Raman-active functional groups in chocolate are the C-C, and C-O-C in the range of ∼ 800 to 970 $cm^{-1}$. Bands at 473 $cm^{-1}$, 636 $cm^{-1}$ and 1299 $cm^{-1}$ were used to identify lactose, sucrose and fat (mostly cocoa butter and milk fat)[1], respectively, while relatively strong bands at 847 $cm^{-1}$ and 869 $cm^{-1}$ are due to the C−O−C characteristic group of sugar.

The Raman spectra revealed a stronger contribution of sucrose than lactose for the non-bloomed solid white chocolate, but an increased contribution of lactose on the surface of liquid white chocolate, and a contribution less than liquid but more than solid in the bloomed state. It can be explained based on the small concentration of lactose on the surface of the solid chocolate, which when melted is spilled onto the surface, resulting in the increase of the lactose signal. For the bloomed state some part of the lactose goes back into the inner region. These features identified the sugar bloom and implied the moisture and lactose as primary cause of the bloom. Sugar bloom may be decreased by decreasing the lactose concentration which may have important industrial applications.

Raman spectra obtained using 785 nm laser in the region of 900 ∼ 1200 $cm^{-1}$ reveal C−C skeletal stretching modes (Fig.2). These spectra have large contributions from the cocoa butter. Solid white chocolate has a sharp peak at 1060 $cm^{-1}$ and a weak shoulder centered at 1089 $cm^{-1}$, while the liquid and bloomed white sample displayed weak peaks at 1064 $cm^{-1}$ and 1061 $cm^{-1}$, but strong peaks at 1079 $cm^{-1}$ and 1084 $cm^{-1}$. The different line widths of several bands in Fig. 2 indicate different crystalline structures of the considered samples.

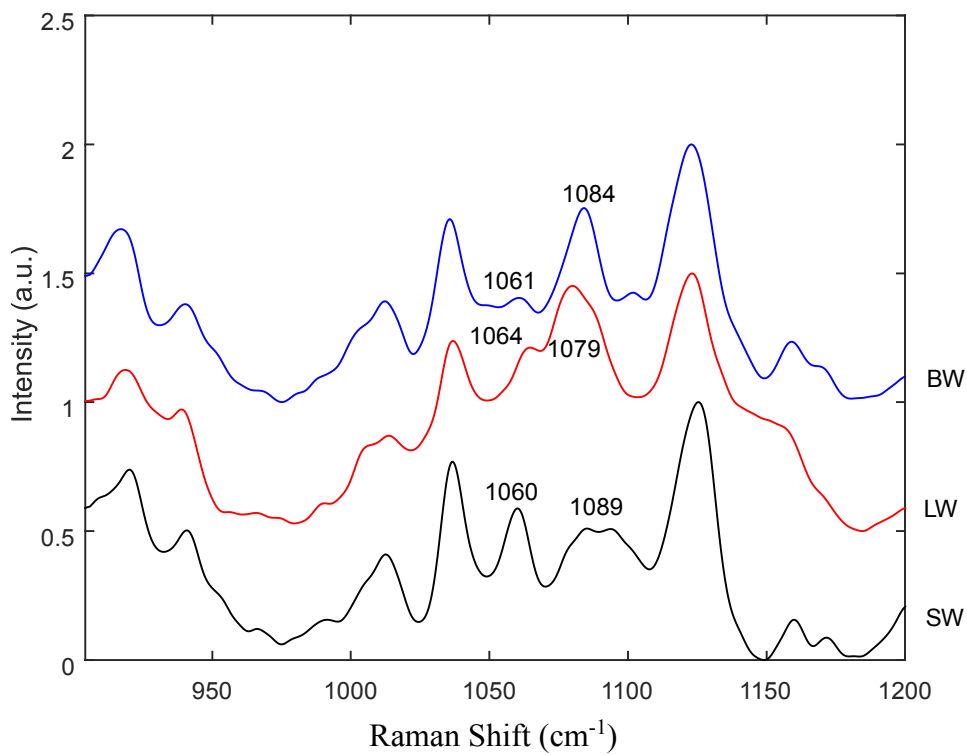

**Figure 2**. Raman spectra of solid (SW), liquid (LW) and bloomed (BW) white chocolate in the 900 ∼ 1200 $cm^{-1}$ range.

Fig. 3 shows Raman spectra of bloomed dark (BD) and bloomed white (BW) chocolate in several spectral ranges. Raman spectra of bloomed dark chocolate were also obtained using 785 nm laser with 60 s acquisition time. The $CH_3$ and $CH_2$ deformation bands at 1438 and 1459 $cm^{-1}$ show significant differences in the spectra of bloomed dark compared to the bloomed white chocolate such as the greater intensity ratio of the 1459 $cm^{-1}$ to 1438 $cm^{-1}$ bands. Similar bands and band ratios in the region of 1050 $cm^{-1}$ - 1200 $cm^{-1}$ identified the VI polymorph in cocoa butter which are also present in the dark chocolate bloom. The region of 848 - 868 $cm^{-1}$ with contributions of lactose and sucrose also displayed similar features because the different lactose concentrations.

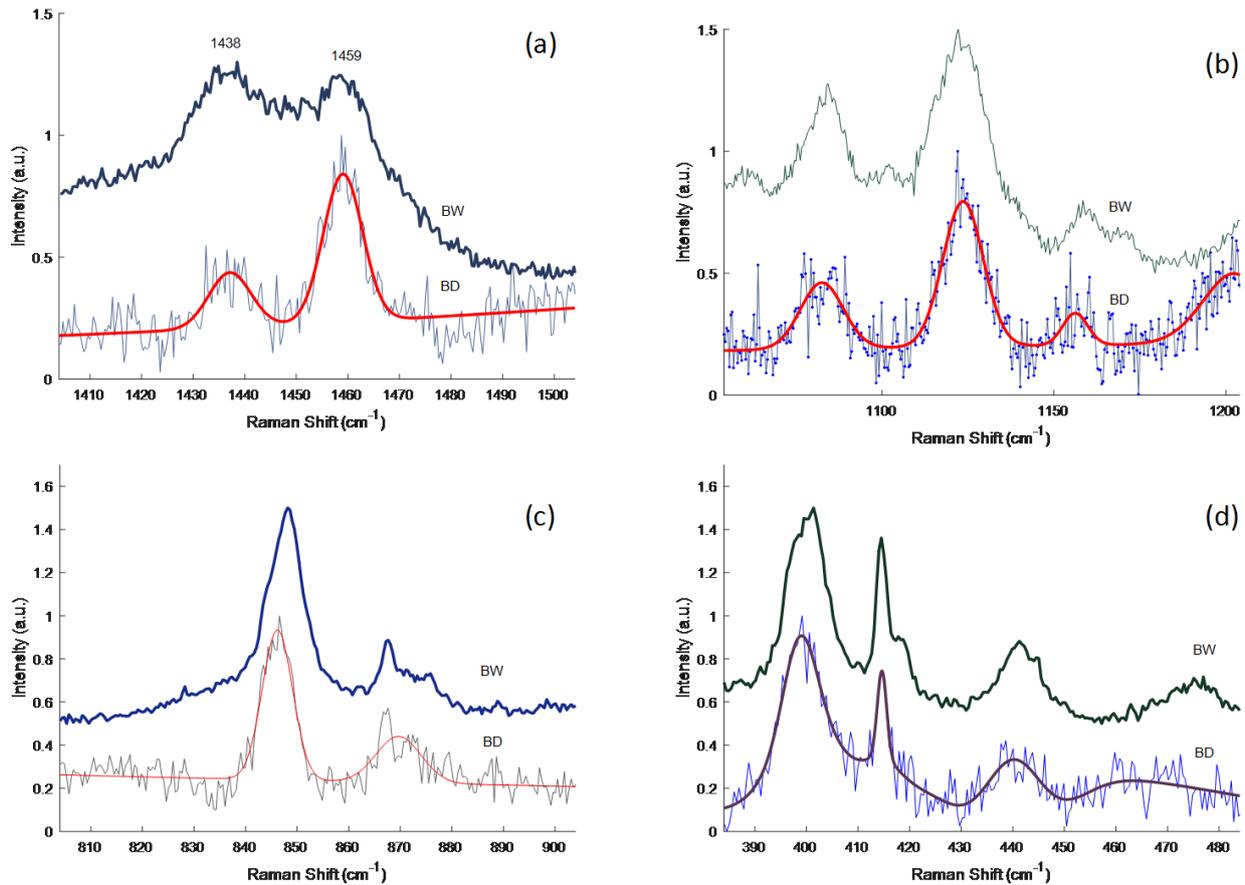

**Figure 3.** Raman spectra of bloomed dark (BD) and bloomed white (BW) chocolate in the (a) 1400 - 1500 $cm^{-1}$, (b) 1050 - 1200 $cm^{-1}$, (c) 800 ~ 900 $cm^{-1}$, and (d) 380 - 480 $cm^{-1}$ ranges.